\documentclass[twocolumn,apj]{emulateapj}

\usepackage{hyperref}
\usepackage{longtable}
\usepackage{natbib}
\usepackage{graphicx}
\usepackage{tabularx}
\usepackage{amsmath,amssymb}
\usepackage{txfonts}
\usepackage{epstopdf}
\usepackage{natbib}
\usepackage{changes}
\usepackage{bm}

\newcolumntype{L}[1]{>{\raggedright\arraybackslash}p{#1}}
\newcolumntype{C}[1]{>{\centering\arraybackslash}p{#1}}
\newcolumntype{R}[1]{>{\raggedleft\arraybackslash}p{#1}}
\newcolumntype{Y}{>{\centering\arraybackslash}X}

\definechangesauthor[color=red]{LF}
\definechangesauthor[color=blue]{PH}
\definechangesauthor[color=green]{LM}
\definechangesauthor[color=lime]{AV}
\definechangesauthor[color=olive]{SL2}
\definechangesauthor[color=violet]{SL}

\newcommand\Bv{\bm{B}}
\newcommand\Jv{\bm{J}}
\newcommand\Ev{\bm{E}}

\newcommand\uv{\bm{u}}
\newcommand\kv{\bm{k}}
\newcommand\di{d_{\rm i}}

\newcommand\rhoi{\rho_{\rm i}}

\begin{document}
\title{Solar wind turbulent cascade from MHD to sub-ion scales: large-size 3D hybrid particle-in-cell simulations}
\author{Luca~Franci}
\affil{Dipartimento di Fisica e Astronomia, Universit\`a di Firenze,
Firenze, Italy.}
\author{Simone~Landi}
\affil{Dipartimento di Fisica e Astronomia, Universit\`a di Firenze,
Firenze, Italy.}
\author{Andrea~Verdini}
\affil{Dipartimento di Fisica e Astronomia, Universit\`a di Firenze,
Firenze, Italy.}
\author{Lorenzo~Matteini}
\affil{Department of Physics, Imperial College London, UK.}
\affil{LESIA, Observatoire de Paris, Meudon, France}
\author{Petr~Hellinger}
\affil{Astronomical Institute, CAS, Prague, Czech Republic}

\date{\today}
 
\begin{abstract}
Spectral properties of the turbulent cascade from fluid to kinetic
scales in collisionless plasmas are investigated by means of
large-size three-dimensional (3D) hybrid (fluid electrons, kinetic
protons) particle-in-cell simulations. Initially isotropic Alfv\'enic
fluctuations rapidly develop a strongly anisotropic turbulent cascade,
mainly in the direction perpendicular to the ambient magnetic
field. The omnidirectional magnetic field spectrum shows a double
power-law behavior over almost two decades in wavenumber, with a
Kolmogorov-like index at large scales, a spectral break around ion
scales, and a steepening at sub-ion scales. Power laws are also
observed in the spectra of the ion bulk velocity, density, and
electric field, both at magnetohydrodynamic (MHD) and at kinetic
scales.  Despite the complex structure, the omnidirectional spectra of
all fields at ion and sub-ion scales are in remarkable quantitative
agreement with those of a two-dimensional (2D) simulation with similar
physical parameters. This provides a partial, a-posteriori validation
of the 2D approximation at kinetic scales. Conversely, at MHD scales,
the spectra of the density and of the velocity (and, consequently, of
the electric field) exhibit differences between the 2D and 3D
cases. Although they can be partly ascribed to the lower spatial
resolution, the main reason is likely the larger importance of
compressible effects in a full geometry. Our findings are also in
remarkable quantitative agreement with solar wind observations.
\end{abstract}


\section{Introduction}

In-situ data from solar and heliospheric spacecraft missions provide
observations of the solar wind plasma and electromagnetic fluctuations
in the frequency range $10^{-5}~\mathrm{Hz}<f<10^2~\mathrm{Hz}$.  The
power spectra of such fluctuations typically exhibit a power-law
behavior over several decades in frequency, with different power-law
indices at scales larger or smaller than about $1\mathrm{Hz}$,
corresponding to the characteristic proton spatial
scales~\citep[e.g.][]{Alexandrova_al_2009,Sahraoui_al_2010,
  Roberts_2010, Chen_2016}. 

Measurements of the third-order structure functions verify that
power-law spectra at scales well above the proton scales (hereafter
MHD scales) result from a turbulent cascade
\citep{MacBride_al_2005, SorrisoValvo_al_2007, MacBride_al_2008}.
Recently, the exact law for the third-order structure function has
been extended to the case of homogeneous incompressible Hall-MHD
turbulence and applied to 2D HPIC simulations
\citep{Hellinger_al_2017b}. Those numerical results suggest that the
cascade likely continues all the way down to sub-proton scales
(hereafter kinetic scales) via the Hall term. Although a direct
evidence is still missing, this is further supported by the fact that
the measured cascade rate in the solar wind 
is consistent with the proton heating rate
\citep{Vasquez_al_2007, Marino_al_2008, Stawarz_al_2009}.  In this
context, direct numerical simulations of turbulent plasmas are not
only useful for interpreting the nature of the solar wind fluctuations
at MHD scales, but also represent a tool to understand how energy is
channeled to protons and electrons.

At MHD scales, the solar wind fluctuations are predominantly
Alfv\'enic: the magnetic field and the ion bulk velocity are observed
to be dominated by their transverse components with respect to the
ambient magnetic field. The former typically shows a Kolmogorov-like
spectrum, with a slope of $\sim-5/3$, while the latter is usually
flatter, with a spectral index closer to $-3/2$.
\citep[e.g.][]{Podesta_al_2007,Salem_al_2009,Wicks_al_2011,Tessein_al_2009}.
The electric field is strongly coupled to the ion bulk velocity
\citep{Chen_al_2011b}.  

When reaching the proton kinetic scales, both the magnetic and
velocity spectra are observed to steepen.  The former has a spectral
index varying between $-4$ and $-2$ at sub-ion scales
\citep[e.~g.]{Leamon_al_1998, Smith_al_2006b,Sahraoui_al_2010}, but
typically close to $\sim-2.8$ in the range between the ion and the
electron scales \citep{Alexandrova_al_2009, Alexandrova_al_2012,
  Sahraoui_al_2013}.  The latter decouples from the magnetic field and
it typically shows a steeper, and much more variable, power law
\citep[e.g.][]{Safrankova_al_2016}. On the contrary, the electric
field spectrum flattens \citep{Bale_al_2005,
  Sahraoui_al_2009, Salem_al_2012}, dominating over the magnetic
field's, with a typical spectral index around $-0.8$
\citep[e.g.][]{Stawarz_al_2016, Matteini_al_2017}, as predicted from 
the generalized Ohm's law \citep{Franci_al_2015b}.  
Moreover, an increase in the magnetic
compressibility \citep{Salem_al_2012,Kiyani_al_2013} and a reduced
variance anisotropy \citep{Podesta_TenBarge_2012} are also observed.

Finally, the spectrum of the density fluctuations is very
peculiar. It exhibits, unlike all the other fields, a sort of triple
power-law behavior, with consequently two different breaks
\citep{Safrankova_al_2015}. Its slope is compatible with $\sim-5/3$ in
the MHD range, close to $-1$ and $\sim-2.8$ at scales slightly larger
or smaller than the ion characteristic scales, respectively
\citep{Chen_al_2012}.

Determining the physical scale(s) and process(es) associated to the
ion-scale break in the magnetic field spectrum is not straightfoward.
The ion inertial length, $\di$, and the ion gyroradius, $\rhoi =\di
\sqrt{\beta}$ ($\beta$ being the ion plasma beta), are very similar
under typical solar wind conditions ($\beta \sim 1$).  Observational
results suggest that the transition likely occurs in correspondence with
the larger between the two when they are well separated
\citep{Chen_al_2014b} or to a combination of the two in the
intermediate-beta regime \citep{Bruno_Trenchi_2014}.  Although the
nature of the power-law behavior of the solar wind fluctuations at
kinetic scales is still under debate, an increasing consensus has
recently emerged about the fundamental role of coherent structures and
magnetic reconnection in shaping the magnetic field spectrum near and
below the ion scales \citep[e.g.][]{Franci_al_2016b,
  Cerri_Califano_2016, Mallet_al_2017, Loureiro_Boldyrev_2017b,
  Franci_al_2017}.

In the last decade, many 2D numerical simulations near and below the
ion scales were able to reproduce some aspects of plasma turbulence
\citep[e.~g.][]{Gary_al_2008, Parashar_al_2009,
  Markovskii_Vasquez_2010, Camporeale_Burgess_2011, Servidio_al_2012,
  Wan_al_2012, Cerri_al_2016, Cerri_Califano_2016}. Despite the
different large-scale initial conditions or forcing, a certain
agreement is observed between their results, suggesting that the
spectral behavior at kinetic scales is quite independent from the
dynamics at MHD scales.

Recently, very high-resolution 2D HPIC simulations fully
covered the transition between the fluid and the kinetic scales. In
particular, \citet{Franci_al_2015a} and \citet{Franci_al_2015b}
produced extended turbulent spectra with well-defined power laws for
the magnetic, ion bulk velocity, density, and electric fluctuations,
in agreement with solar wind observations.  Moreover,
\citet{Franci_al_2016b} unambiguously determined the ion-scale
break in the magnetic field spectrum and recovered the observed
dependence on the ion characteristic scales. 

Although 2D HPIC simulations allow for simultaneously covering three
decades in wavenumber across the ion scales, they imply limitations
for the turbulent dynamics and for the onset of plasma
instabilities. This is particularly relevant for the solar wind plasma, in
which the spherical expansion of the mean flow on the one hand shapes
the turbulent anisotropy at MHD scale \citep{Dong_al_2014,
  Verdini_Grappin_2015b, Verdini_Grappin_2016}, and on the other hand
continuosly drives instabilities of proton velocity distribution
functions at kinetic scales \citep{Hellinger_al_2015,
  Hellinger_al_2017a}.

In recent years, an increasing
number of 3D numerical simulations have investigated the development and the
properties of the turbulent cascade around and below the ion scales,
employing different methods and models
\citep[e.g.][]{Shaikh_Shukla_2009, Chang_al_2011, Howes_al_2011,
  Gary_al_2012, Boldyrev_al_2013, Gomez_al_2013, Meyrand_Galtier_2013,
  Rodriguez_al_2013, TenBarge_al_2013a, Passot_al_2014, Vasquez2015,
  Servidio_al_2015, Wan_al_2015, Wan_al_2016, Valentini_al_2017,
  Cerri_al_2017b}.
However, to our knowledge, 3D kinetic simulations have not been
accurate enough yet to obtain clear and extended power laws for the
electromagnetic and plasma fluctuations, spanning both the MHD and the
kinetic range, consistently with solar wind observations. Moreover, while
the different behavior of intermittency and dissipation between 2D and 3D 
has been investigated \citep{Wan_al_2016}, a
quantitative comparison of the spectral properties for all
fields (e.g., spectral indices, scale of the break) in the two cases
is still lacking in the literature.

In this work, we extend our 2D numerical studies, investigating the
physics behind the transition from MHD to kinetic scales in a full 3D
geometry, which allows for a more realistic representation of the
solar wind turbulent plasma.  Via a quantitative analysis of the
spectral properties, we test the limitations of the reduced geometry
and validate the results of our previous 2D HPIC simulations.

The paper is organized as follows. In Sec.~\ref{sec:setup}, we
introduce the numerical and physical setup, along with the definitions
of spectra. In Sec.~\ref{sec:results}, we describe our numerical
results, first focusing on the development of the turbulent cascade
(Sec.~\ref{subsec:evolution}) and later on the spectral properties of
the quasi-stationary fully-developed turbulent state
(Sec.~\ref{subsec:developed}). In Sec.~\ref{subsec:comparison}, we
provide a direct quantitative comparison of our 3D results with
those of a previous 2D simulation of comparable size and similar
physical parameters.  Finally, in Sec.~\ref{sec:conclusions}, we
summarize and discuss our findings. 

\section{Numerical setup and initial conditions}
\label{sec:setup}

We employ the hybrid particle-in-cell code CAMELIA (Current Advance
Method Et cycLIc leApfrog), where the electrons are considered as a
massless, charge neutralizing fluid, whereas the ions (protons) are
described by a particle-in-cell model and are advanced by the Boris
scheme (see \citealt{Matthews_1994} for detailed model equations).

The characteristic spatial and temporal units in this model are
the ion inertial length $\di=v_A/\Omega_{\rm i}$, $v_A$ being
the Alfv\'en speed, and the inverse ion gyrofrequency $\Omega_{\rm
  i}^{-1}$, respectively.

We ran a simulation employing a periodic cubic grid with spatial
resolution $0.25 \, \di$, $512^3$ grid
points, box size $L_{\rm box} = 128\,\di$, and $2048$
particle-per-cell (ppc) representing protons.  The resistive
coefficient is set to the value $\eta=1.5 \times 10^{-3}~4\pi
v_Ac^{-1}\Omega_{\rm i}^{-1}$, to prevent the accumulation of magnetic
energy at the smallest scales. The ions are advanced with a time step
$\Delta t=0.05 \, \Omega_{\rm i}^{-1}$, while the magnetic field $\Bv$
is advanced with a smaller time step, $\Delta t_B = \Delta t/10$.

We assume a uniform magnetic field directed along the
$z$-direction, $\Bv_0=B_0\hat{\mathbf{z}} = 1$. Fields and wavevectors
are always defined as parallel ($\parallel$) and perpendicular ($\perp$) with
respect to $\Bv_0$.  We also assume a uniform density, equal for ions
and electrons, $n_{\rm i}=n_{\rm e}=n$. Both species have uniform and
isotropic temperatures, $T_{\rm i}$ and $T_{\rm e}$, such that
$\beta_{\rm i}=\beta_{\rm e}=0.5$ (being $\beta_{\rm i,e}=8\pi n
K_{\rm B} T_{\rm i,e} / B^2_0$ the plasma betas and $K_{\rm B}$ the
Boltzmann's constant).

We initialize the simulation by imposing linearly polarized shear
Alfv\'enic fluctuations with random phases, i.e., the fluctuations are
perpendicular to the plane defined by their wavevector and the mean
field. In this way, the initial kinetic and magnetic fluctuations are
almost at equipartition (within $10\%$) and have vanishing
correlation. The velocity fluctuations are divergence-less and the
density fluctuations are vanishing (in the limit of numerical noise).
Fourier modes of equal amplitude are excited in the range $k_0 < k \, \di <
k_{\mathrm{inj}}$, where $k = \sqrt{k_x^2 + k_y^2 +
  k_z^2}$. The minimum wavenumber is $k_0 = 2 \pi / L_{\rm box} \simeq 0.05 \, \di^{-1}$, while the maximum injection scale is
$k_{\mathrm{inj}} \simeq 0.25 \, \di^{-1}$.  

We define the 3D axisymmetric spectrum of a generic field $\bm{\Psi}$ as the energy
of its 3D Fourier modes averaged over rings delimited by $k_\bot$ and
$k_\bot + dk_\bot$ (being $k_\bot$ the perpendicular wavenumber and
$dk_\bot = k_0$),
\begin{equation}
P_{\mathrm{3D}}(k_\bot,k_\parallel)=\frac{1}{k_\bot}\sum_{\sqrt{k_x^2+k_y^2} = k_\bot}\hat{\Psi}^2_{\mathrm{3D}}(k_x,k_y,k_z),
\label{def:spectrum3D}
\end{equation}
where $\hat{\Psi}$ are $\bm{\Psi}$'s Fourier coefficients.
This is statistically representative of a random 2D slice of the 3D
Fourier space in the direction of $\kv_\parallel$.  

The 2D spectrum is obtained by integrating the energy of the Fourier 
modes contained in the above defined rings,
\begin{equation}
\begin{aligned}
P_{\mathrm{2D}}(k_\bot,k_\parallel) &= \sum_{\sqrt{k_x^2+k_y^2} =
  k_\bot}\hat{\Psi}^2_{\mathrm{3D}}(k_x,k_y,k_z) = \\ &= k_\bot \,
P_{\mathrm{3D}}(k_\bot,k_\parallel).
\end{aligned}
\label{def:spectrum2D}
\end{equation}
A further integration yields the onedimensional (1D) reduced 
perpendicular and parallel spectra, 
\begin{equation}
P_{\mathrm{1D},\bot} (k_\bot) = \sum_{k_\parallel} P_{\mathrm{2D}}^\Psi(k_\bot,k_\parallel),
\label{def:spectrum1Dper}
\end{equation}
\begin{equation}
P_{\mathrm{1D},\parallel} (k_\parallel) = \sum_{k_\bot} P_{\mathrm{2D}}^\Psi(k_\bot,k_\parallel).
\label{def:spectrum1Dpar}
\end{equation}

The 1D omnidirectional spectrum can be obtained by integrating the
energy of the Fourier modes over spherical shells delimited by $k$ and
$k + dk$,
\begin{equation}
P_\mathrm{1D}(k)=\sum_{\sqrt{k_x^2+k_y^2+k_z^2}=k}\hat{\Psi}^2_{\mathrm{3D}}(k_x,k_y,k_z).
\label{def:spectrum1Diso}
\end{equation}

Finally, we define the root mean square value (rms) as
\begin{equation}
 \Psi^{\rm rms}=\sqrt{\langle \Psi^2\rangle - \langle \Psi \rangle^2},
\end{equation}
where $\langle...\rangle$ stands for the real-space average over the
whole simulation domain.  

With these definitions, the initial conditions have
$P^{\mathrm{1D}}_{\Bv} \sim P^{\mathrm{1D}}_{\uv} \propto k^2$, with
$\Bv^{\rm rms}/B_0 \sim 0.4$.  Since almost all the energy is
concentrated at $k\di \sim 0.25$, the estimated non-linear time at the
beginning of the simulation is $ t_{\rm NL} \sim 10~\Omega_{\rm
  i}^{-1}$.

\begin{figure}[t]
\begin{center}
\includegraphics[width=\linewidth]{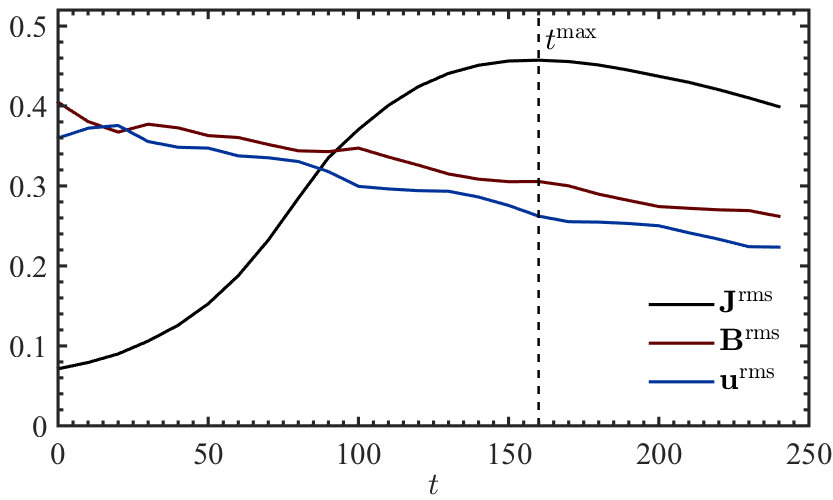}
\caption{Time evolution of the rms of the current density, $\Jv$
  (black line), of the magnetic field, $\Bv$ (red), and ion bulk
  velocity, $\uv$ (blue). The dashed vertical line marks the time of
  the maximum of $\Jv^{\rm rms}$, $t_{\rm max}=160~\Omega_{\rm
    i}^{-1}$.}
\label{fig:evolution}
\end{center}
\end{figure}

\section{Results}
\label{sec:results}

\subsection{Time evolution and development of a turbulent cascade}
\label{subsec:evolution}

\begin{figure*}[t]
\begin{center}
\includegraphics[trim={0 1.9cm 0 0},clip,width=0.40\linewidth]{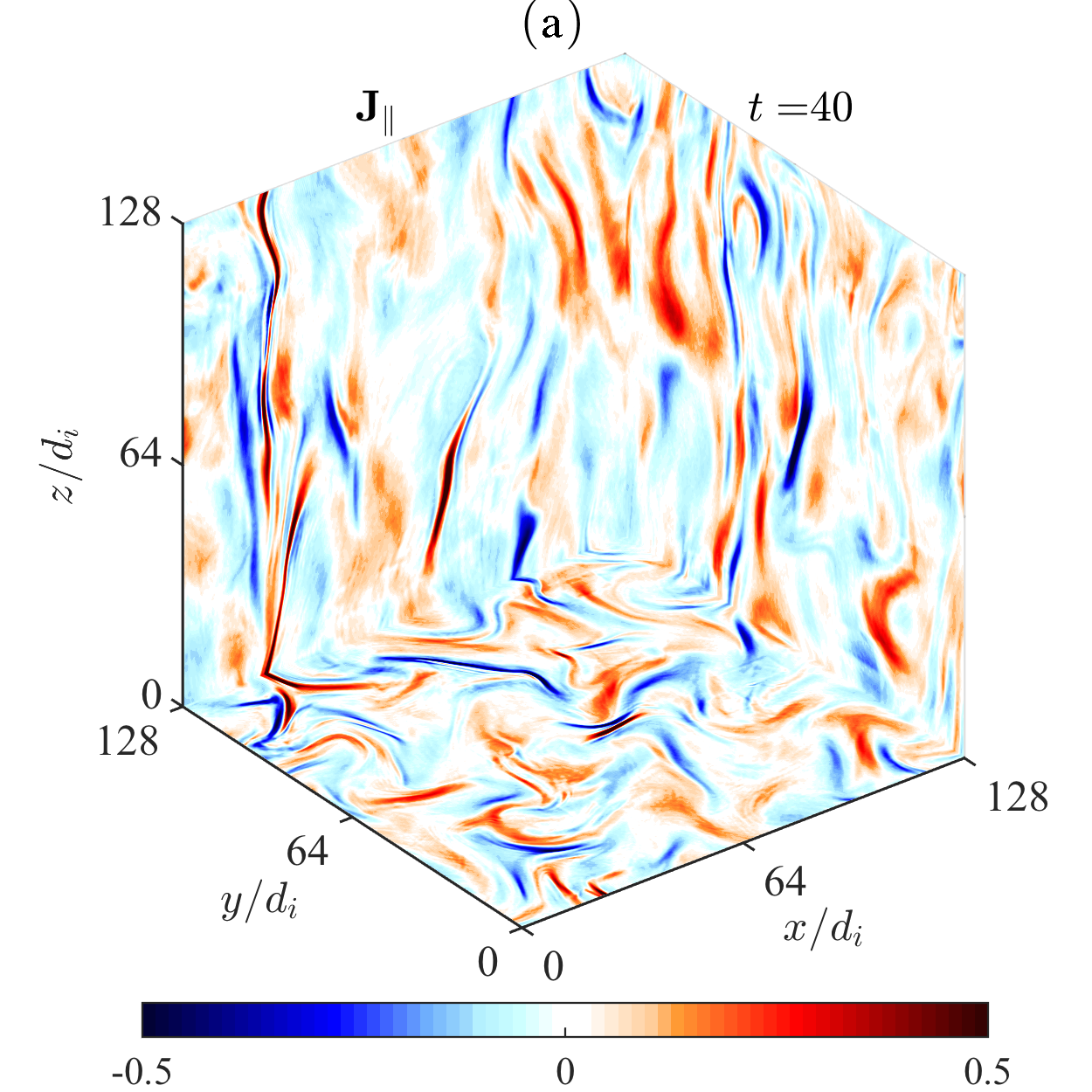}
\includegraphics[trim={0 1.9cm 0 0},clip,width=0.40\linewidth]{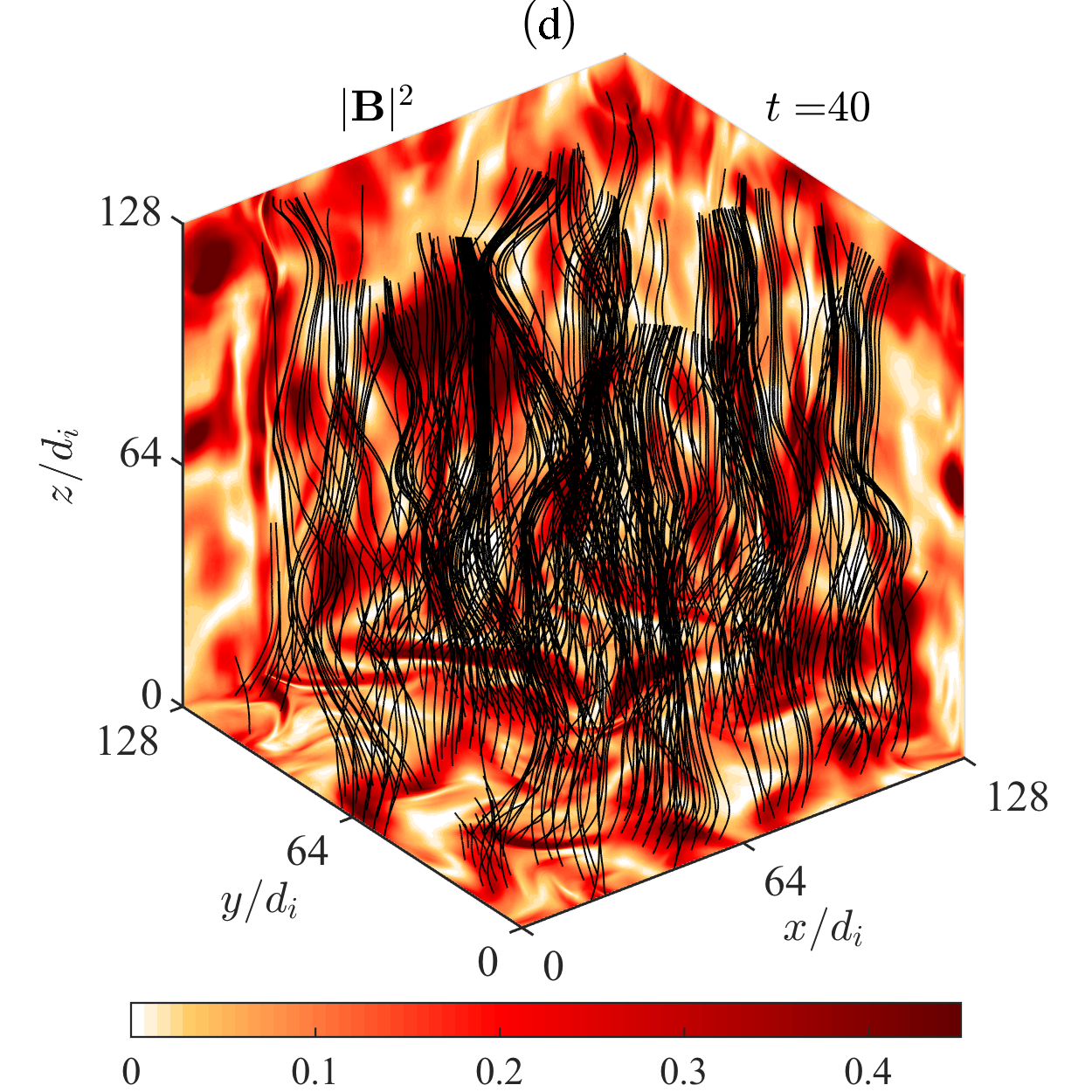}\\
\includegraphics[trim={0 1.9cm 0 0},clip,width=0.40\linewidth]{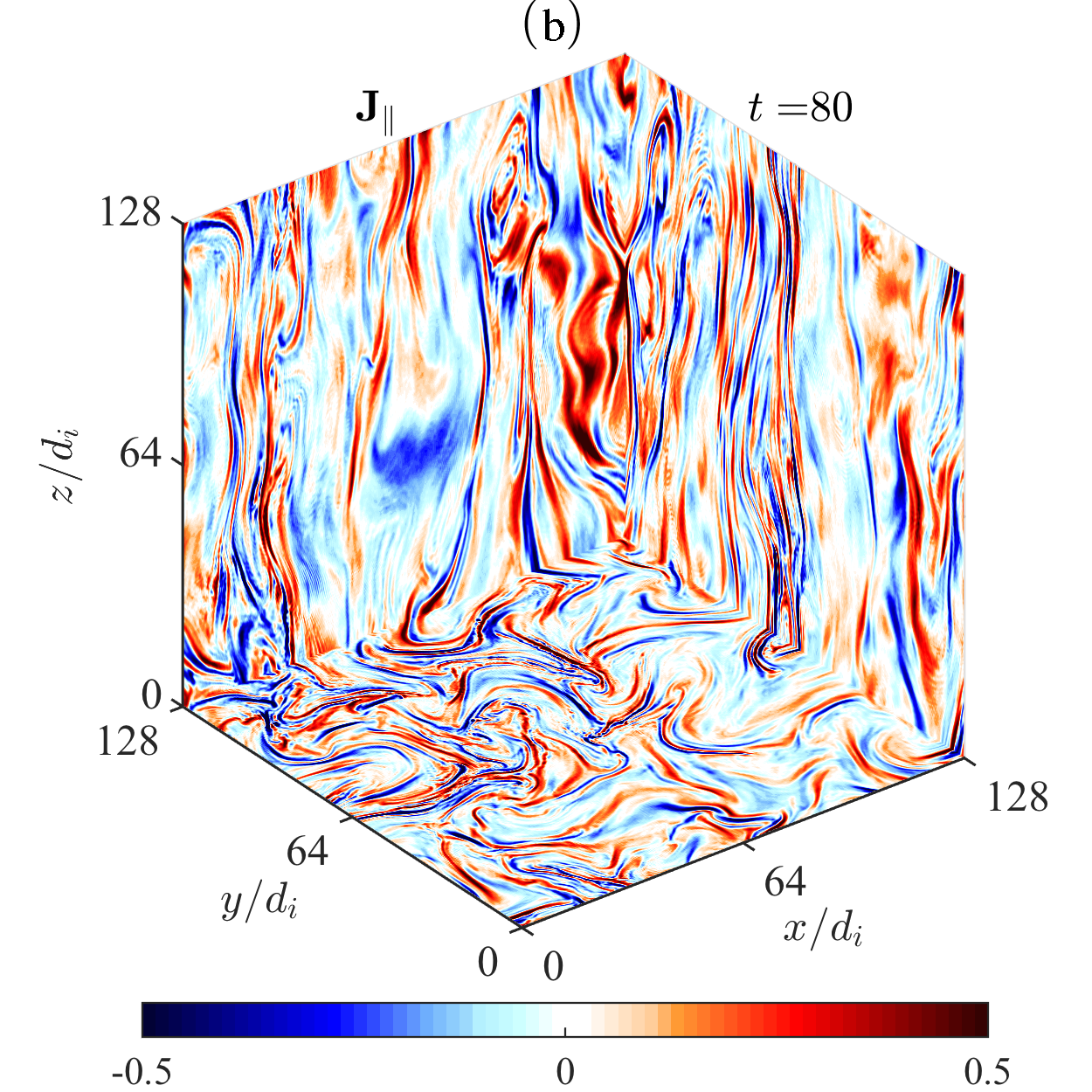}
\includegraphics[trim={0 1.9cm 0 0},clip,width=0.40\linewidth]{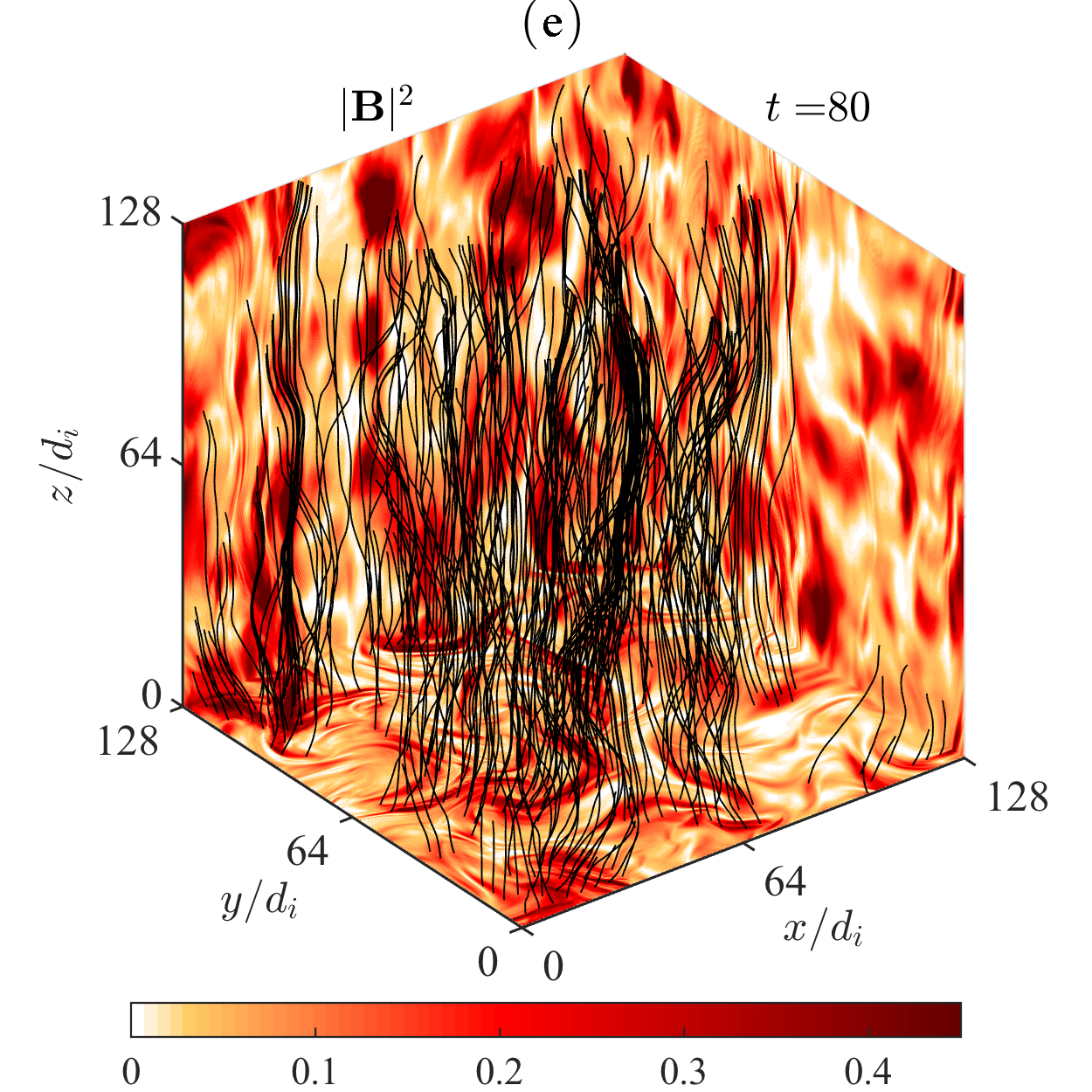}\\
\includegraphics[width=0.40\linewidth]{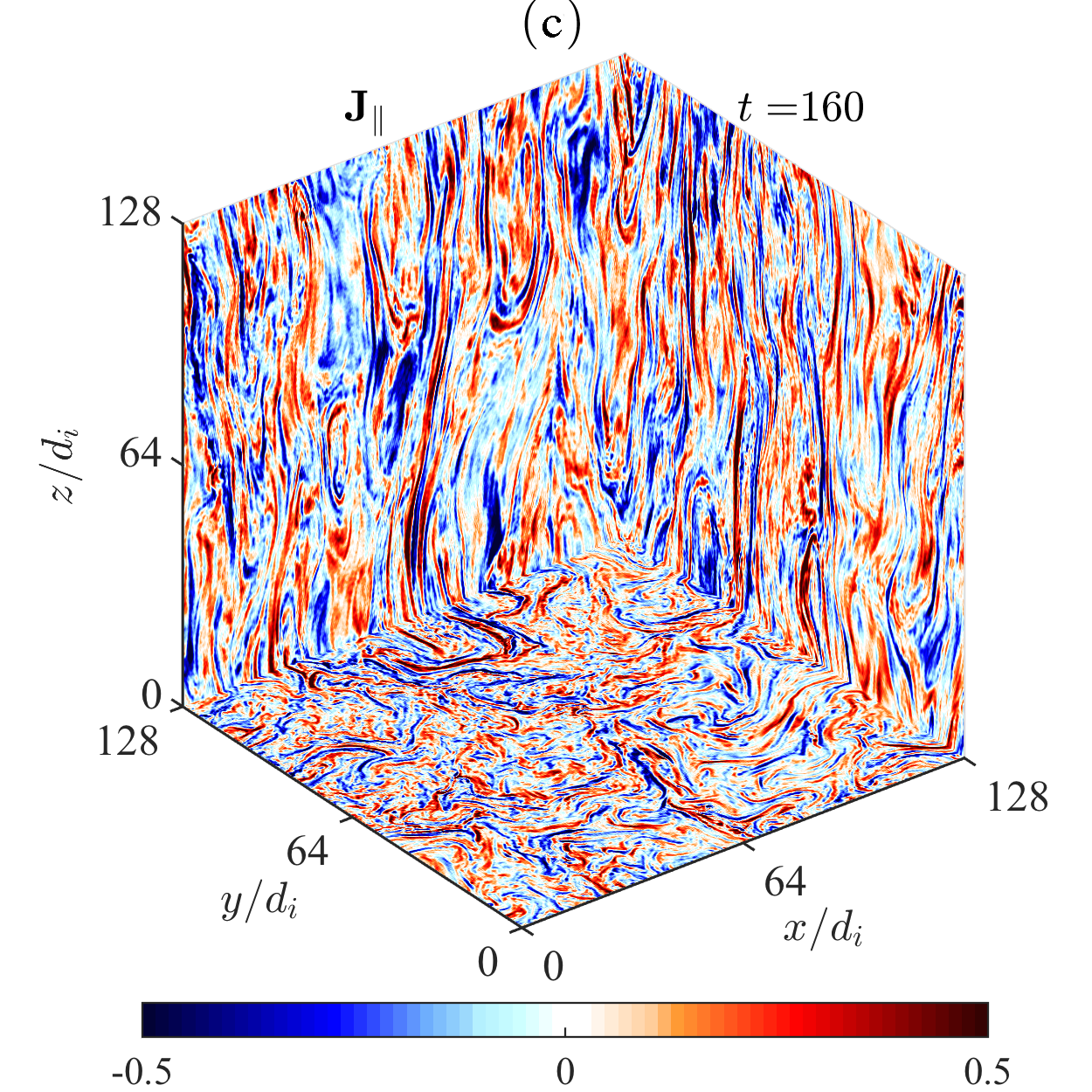}
\includegraphics[width=0.40\linewidth]{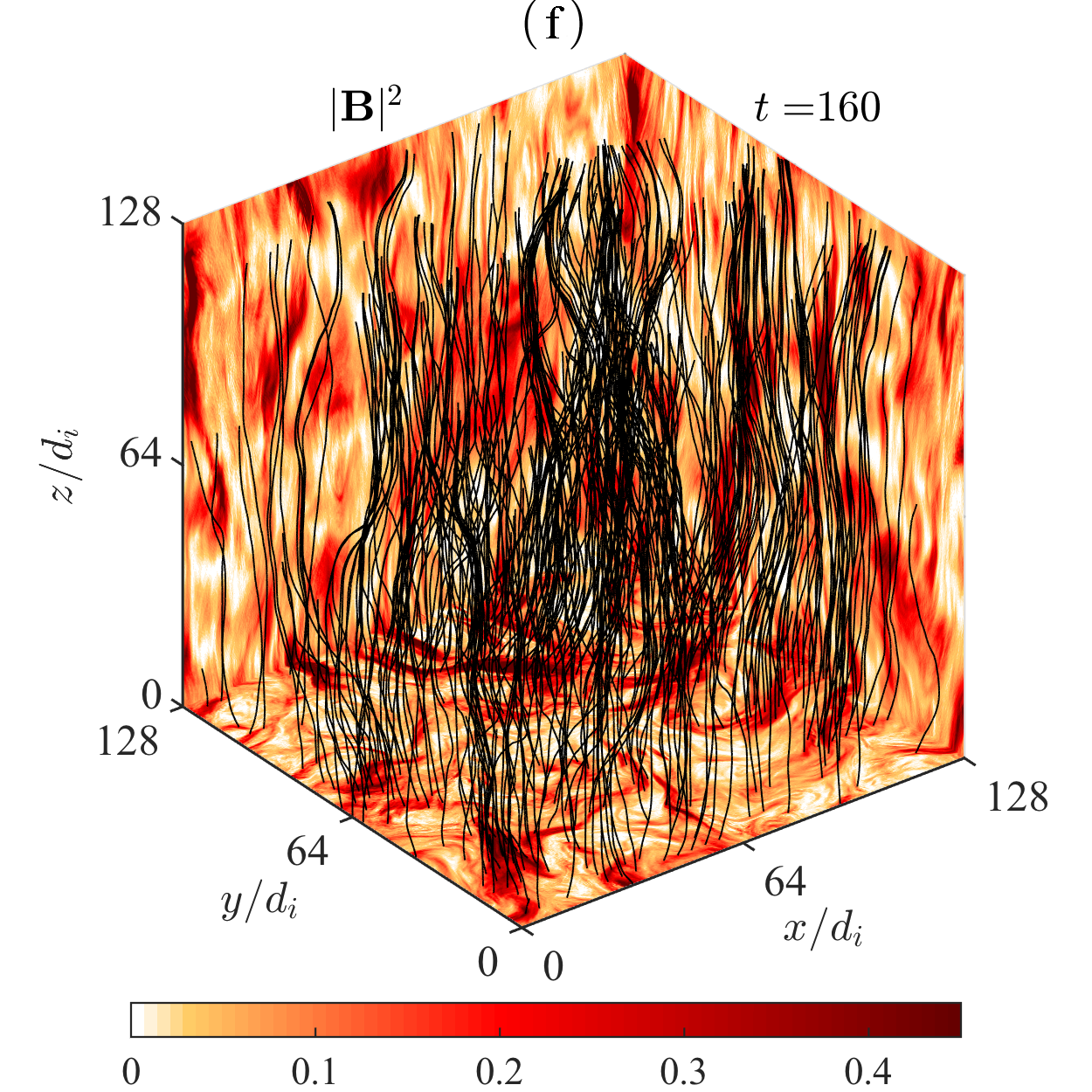}
\caption{3D pseudocolor plots of $\Jv_\parallel$ (panels (a)-(c)) and
  of $|\Bv|^2$ (panels (d)-(f)), with magnetic field lines drawn in
  black. Simulations times are (from top to bottom): $t=40$, $80$, and
  $t=160\,\Omega_{\rm i}^{-1}$.}
\label{fig:3Dcubes}
\end{center}
\end{figure*}
In Fig.~\ref{fig:evolution}, we report the time
evolution of the rms of the current density,
$\Jv^{\rm rms}$ (black line), of the magnetic field, 
$\Bv^{\rm rms}$ (red), and of the ion bulk velocity,
$\uv^{\rm rms}$ (blue line).

$\Jv^{\rm rms}$ increases quite rapidly, until a
maximum is reached at about $t = t_{\rm max} \equiv160~\Omega_{\rm i}^{-1}$ 
(marked with a vertical dashed line) and then it declines
smoothly and slowly. The maximum correspondes to about $15~t_{\rm
  NL}$, which is of the same order as what observed in previous 2D
simulations using a very similar setup \citep{Franci_al_2015b}. Being
the peak in $\Jv^{\rm rms}$ tipically regarded as an indicator of the
maximum turbulent activity \citep{Mininni_Pouquet_2009}, the analysis
of the spectral properties in Sec.\ref{subsec:developed}
will be performed at $t=t^{\rm max}$. Moreover, the
turbulent activity is observed to be quite stable afterwards, so that
all the considerations remain valid until the end of the simulation
($t=240~\Omega_{\rm i}^{-1}$). Both $\Bv^{\rm rms}$ and 
$\uv^{\rm rms}$ decline quasi steadily all
over the simulation, with an excess of magnetic over kinetic
energy of about $10-15\%$ maintained 
throughout the whole evolution.

In Fig.~\ref{fig:3Dcubes}, we report 3D pseudocolor plots of the
current (panels (a)-(c)) and magnetic field structures (panels
(d)-(f)) at three different times, i.e., $t = 40$, $80$, and
$t=160~\Omega_{\rm i}^{-1}=t^{\rm max}$.  Initially, we observe the
formation of intense current sheets having a quasi-2D shape, with a
length in the direction of the mean magnetic field which is of the
order of the box size, a slightly smaller width and a much smaller
thickness, of the order of the ion inertial length (panel (a)).  Later,
the number of current structures increases and some of them are
disrupted, likely because of magnetic reconnection, although clear
signatures of such events are not easily identifiable by eye (b).  At
the time of maximum turbulent activity, the current structures are
much more complex and more uniformly distributed all over the physical
domain, still being characterized by an elongated shape along the $z$
direction (c). 
Correspondingly, the intensity of the magnetic field is shown in
panels (d)-(f), along with magnetic field lines (black lines).
Large-scale intense magnetic structure are initially isotropic, while
they appear more and more filementary and twisted at later times, with
strong gradients in the perpendicular plane and lenght-scale of a few
fraction of the box size in the $z$-direction.

This is also seen in the magnetic field lines, which at early times are
strongly perturbated along $\Bv_0$ and clustered in the perpendicular
plane. Initially, they are modulated by long-wavelength fluctuations
in all directions.  At later times, as the fluctuation amplitude
decreases, field lines have approximately the same parallel wavelength
and their distribution in the perpendicular plane is disordered,
because of the many small-scale structures formed in the $(x,y)$
plane.  This indicates that a large spectral anisotropy has been
developed from the isotropic initial conditions.

The characteristic lenght scale of the parallel modulation remains
equal to approximately $L_{\mathrm{box}}/8$ at $t \gtrsim 80$, so
we will choose this value as an averaging scale in
sec.~\ref{subsec:comparison}, when comparing the real-space structures
of the present 3D simulation with the ones of a previous 2D run.

\begin{figure}[t]
\begin{center}
\includegraphics[width=\linewidth]{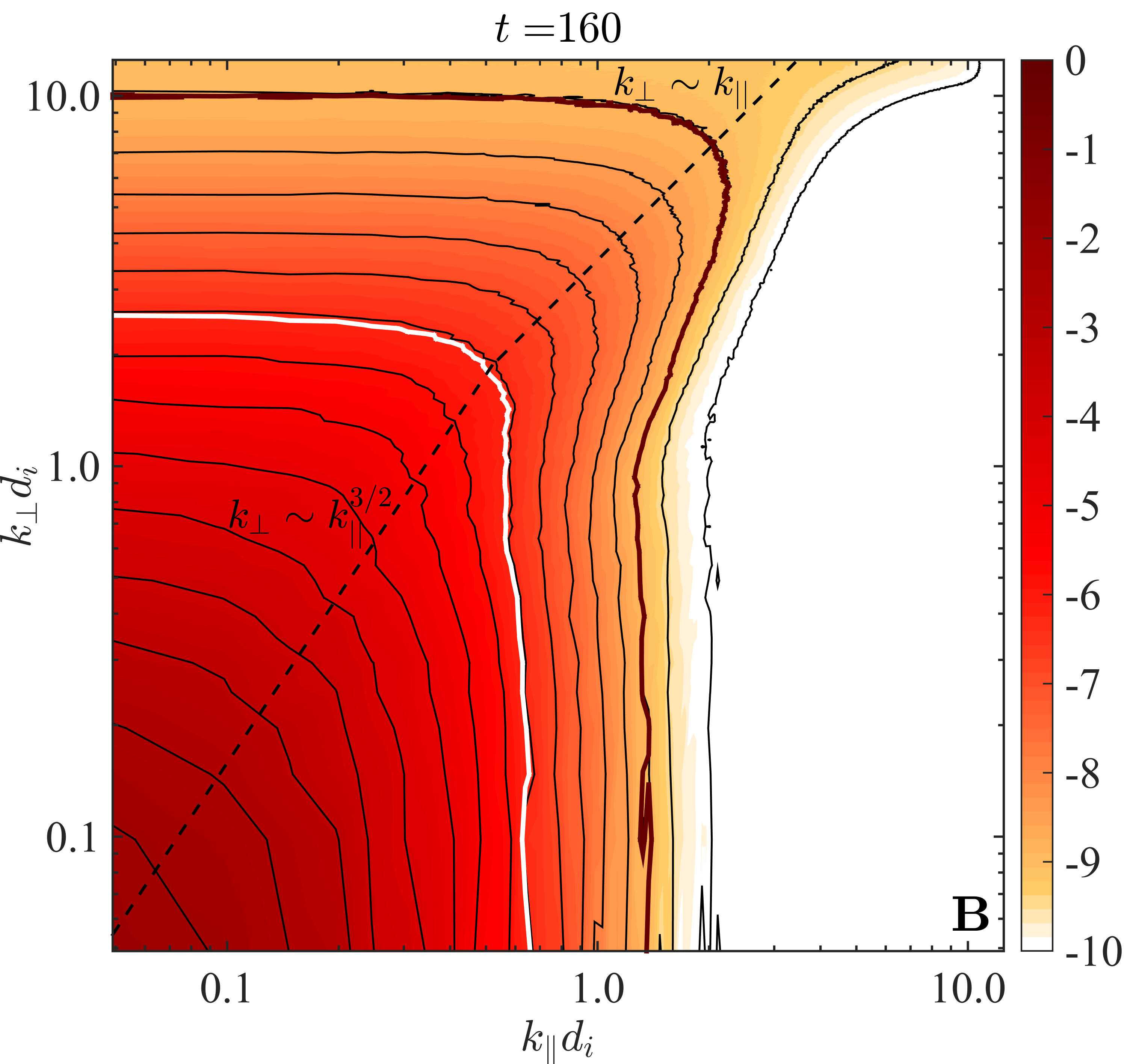}\\
\caption{Isocontours of the 3D spectrum of the magnetic fluctuations,
  $P^{\Bv}_{\mathrm{3D}}$, in the $(k_\parallel,k_\perp)$ space, at
  $t=t^{\rm max}$.}
\label{fig:3Dspectrum}
\end{center}
\end{figure}

\subsection{Fully-developed quasi-stationary state}
\label{subsec:developed}

\begin{figure}[t]
\begin{center}
\includegraphics[width=\linewidth]{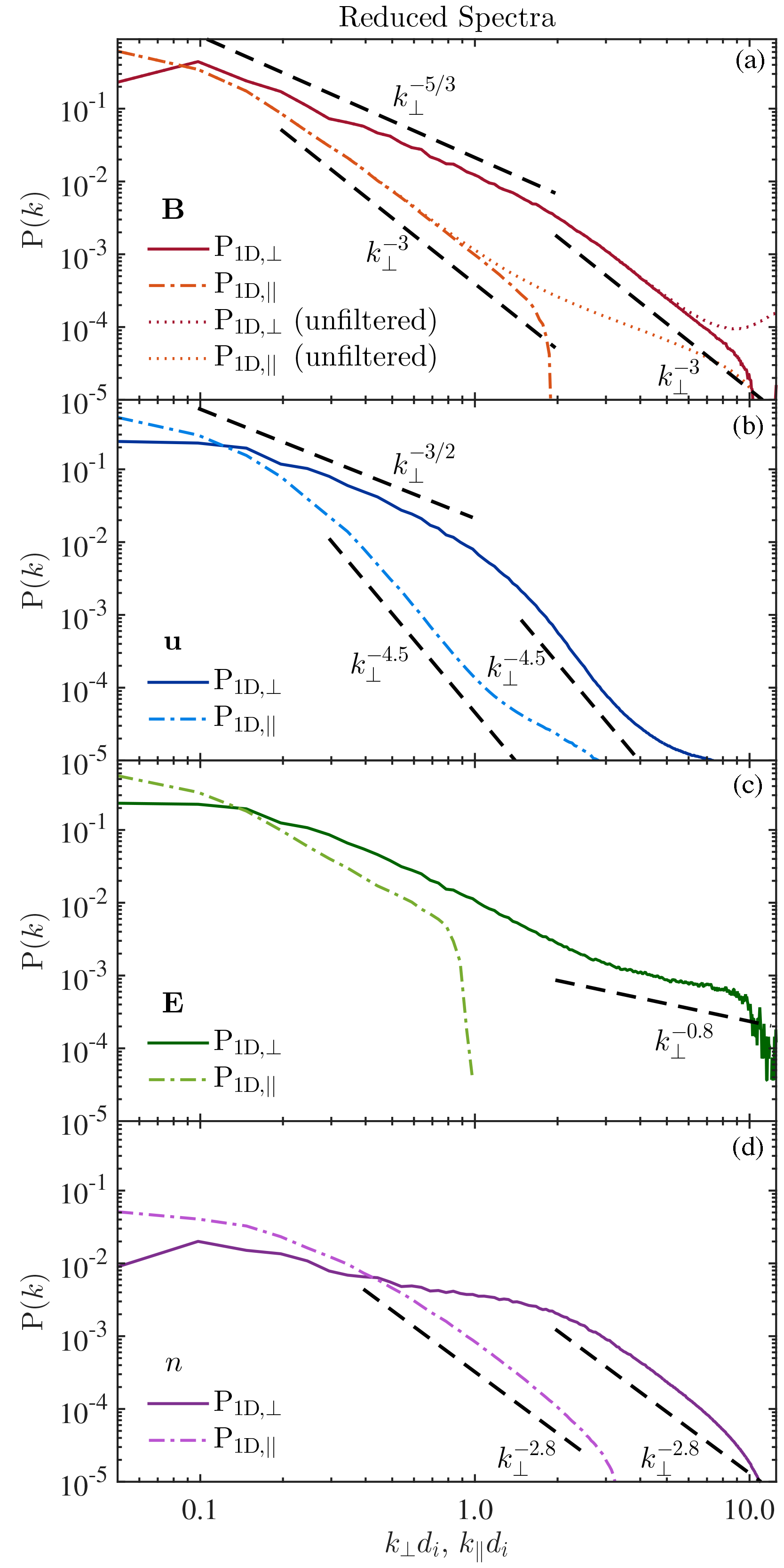}
\caption{1D filtered reduced perpendicular and parallel spectra of
  magnetic (panel (a)), ion bulk velocity (b), electric (c), and
  density (d) fluctuations.  Additionally, the unfiltered spectra are
  shown for the magnetic fluctuations with dotted lines for
  comparison. Characteristic power laws are drawn in black dashed
  lines as a reference.}
\label{fig:1DReducedSpectra}
\end{center}
\end{figure}

The spectral properties of the magnetic fluctuations at $t=t^{\rm
  max}$ are shown in Fig.~\ref{fig:3Dspectrum} by the isocontours of
their 3D axisymmetric spectrum, $P_{\mathrm{3D}}$
(Eq.~\ref{def:spectrum3D}).  These confirm the qualitative behavior
seen in Fig.~\ref{fig:3Dcubes}, i.e., a strong anisotropy develops,
with energy cascading mainly in perpendicular wavenumbers.  A white
isoncontour, corresponding to $k_\bot \di \sim 2.5$, separates the MHD
range from the kinetic range (as will be inferred from
Fig.~\ref{fig:1DReducedSpectra}).  
The different spacing of the isolevels in the perpendicular wavenumber
allows for a rough qualitative evaluation of the spectral index, which
is flatter in the MHD range than in the kinetic one.  In the parallel
direction, instead, no clear power-law behavior is seen at scales
corresponding to the MHD range and the transition between the two 
ranges occurs at smaller wavenumbers.

Note that the spectral anisotropy is scale-dependent in the MHD range,
while it is scale-independent in the kinetic range, as suggested by
the vertices of the isolevels being aligned approximately with the
reference scaling (black dashed lines), $k_\parallel\sim k_\bot^{2/3}$
and $k_\parallel\sim k_\bot$, respectively.

The temporal evolution of the 3D spectrum (not shown) reveals that the
cascading energy reaches the maximum wavenumber (corresponding to the
grid size) first in the perpendicular direction and then spreads in
the parallel direction for high values of $k_\bot$. As a result, an
unphysical accumulation occurs at $k_\bot \di \gtrsim 10$ for all
parallel wavenumbers. When integrating to obtain the 1D reduced
perpendicular and parallel spectra and the 1D omnidirectional spectra,
we will therefore remove this unphysical excitation by putting to zero
the power corresponding to all the isocontours that extend beyond the
dark red line in the 3D spectrum.


In Fig.~\ref{fig:1DReducedSpectra}, we show the 1D reduced 
perpendicular and parallel spectra, 
(Eq.~\ref{def:spectrum1Dper} and ~\ref{def:spectrum1Dpar},
respectively) for the magnetic (panel (a)), ion bulk velocity (b),
electric (c) and density (d) fluctuations.  The reduced spectra
of $\Bv$ 
obtained by directly integrating $P^{\Bv}_{\mathrm{2D}}$
(Eq.~\ref{def:spectrum2D}) are drawn as dotted lines and 
labeled as ``unfiltered''.
At all scales,
the perpendicular transfer of magnetic energy with respect to the 
mean field is always dominant with respect to
the parallel one, reflecting the strong spectral anisotropy already 
observed in Fig.~\ref{fig:3Dspectrum}. In particular,
$P_{\mathrm{1D},\bot}^{\Bv}$ exhibits two different power laws
spanning almost two decades in wavevector: a spectral index
of $-5/3$ at MHD scales, a transition around ion scales, and a
steeper slope $\sim-3$ at sub-ion scales (two reference
power laws are drawn in black dashed lines). The parallel spectrum,
$P_{\mathrm{1D},\parallel}^{\Bv}$, is much steeper at small
wavevectors and then flattens for $k_\parallel \, \di \gtrsim
1$.  

As already mentioned when discussing Fig.~\ref{fig:3Dspectrum}, a
significant energy accumulation, due to numerical effects and not
sufficiently removed by the explicit resistivity, occurs for large
perpendicular wavevectors and it spreads out at all scales in the
parallel direction. As a consequence, noise can significantly affect
$P_{\mathrm{1D},\parallel}^{\Bv}$, causing the flattening observed for
$k_\parallel \, \di \gtrsim 1$.
We estimate a filtering scale, $k_{\rm filter} \di$, at which this
effect unphysically alters the spectral behavior and we filter out all
the isocontours of $P_{\mathrm{2D}}$ beyond that threshold (shown for
the magnetic field as a white line in
Fig.~\ref{fig:3Dspectrum}). Specifically, we choose $k_{\rm filter}
\di = 10$, which corresponds to the perpendicular wavevector where
$P_{\mathrm{1D},\bot}^{\Bv}$ starts to artificially increase. By
testing different values for the threshold, we have checked that the
filtering procedure does not affect the scales larger than
$k_{\rm filter}$, while it removes artificial features only due to the
power accumulation (especially in the parallel spectra).

The smallest scales are problematic for the other fields as well, due
to the finite number of particles (affecting the density and velocity)
and to the computation of finite-differences derivatives of
numerically-affected fields (for the electric field).  We thus repeat
the same filtering procedure for all the fields, setting the
same filtering scale as for $\Bv$.

The filtered reduced spectra of magnetic fluctuations are drawn in
Fig.~\ref{fig:1DReducedSpectra}(a) with solid lines. No flattening is
now seen in $P_{\mathrm{1D},\parallel}^{\Bv}$.  In the range
$0.2\lesssim k_\parallel d_{\rm i}\lesssim 2$, it exhibits a power-law
with a slope very close to that of $P_{\mathrm{1D},\bot}^{\Bv}$ for
$1\lesssim k_\bot \di\lesssim 10$.

\begin{figure*}[t]
\begin{center}
\includegraphics[width=0.8\linewidth]{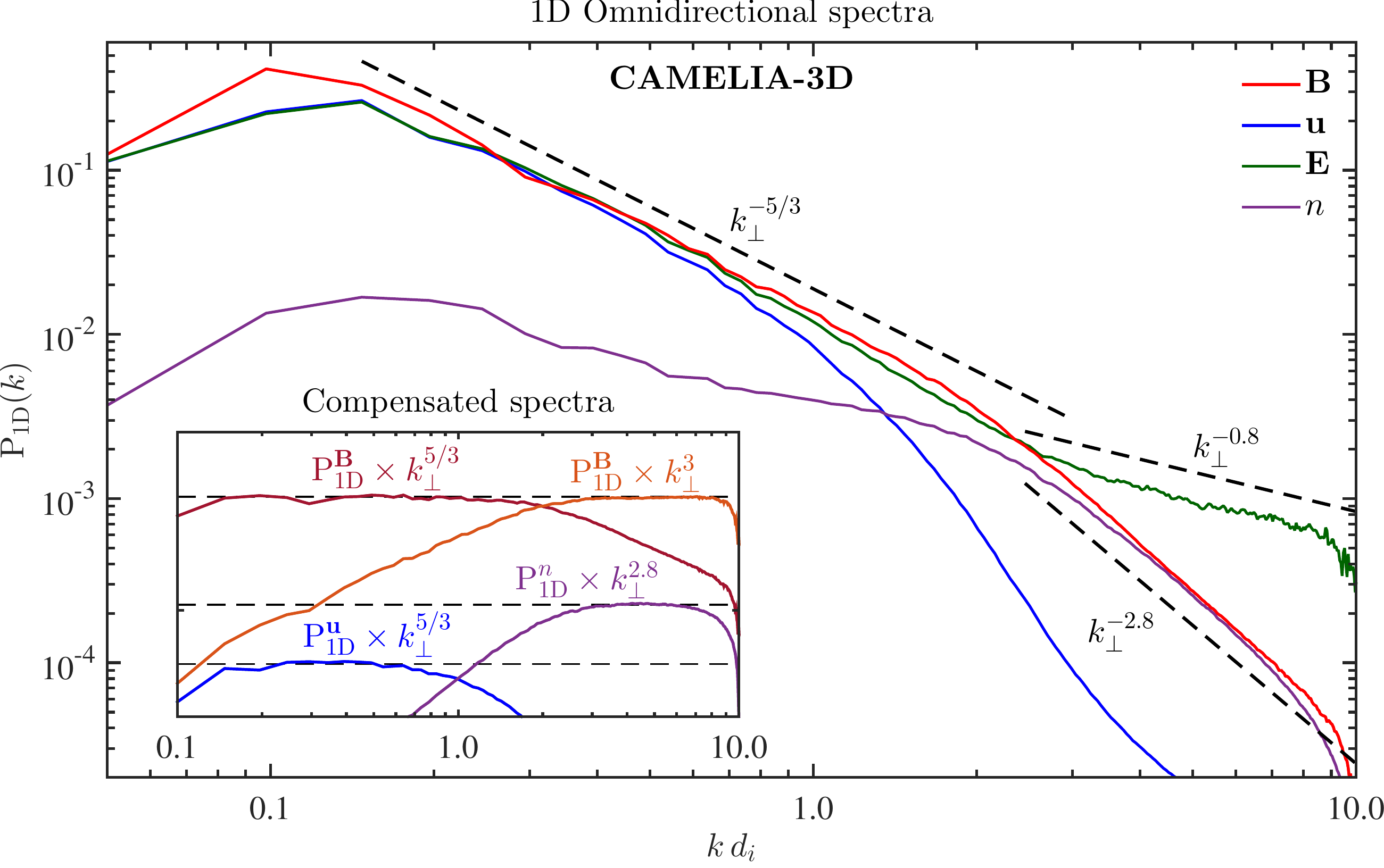}
\caption{1D omnidirectional spectra of magnetic (red), ion bulk velocity
  (blue), electric (green), and density (purple)
  fluctuations. Characteristic power laws are also drawn with dashed
  black lines as a reference. The inset reports the spectrum of
  the magnetic field fluctuations compensated by $k^{5/3}$ (dark red)
  and by $k^{2.9}$ (orange).}
\label{fig:1DOverview}
\end{center}
\end{figure*}

The reduced perpendicular spectrum of the ion bulk velocity,
$P_{\mathrm{1D},\bot}^{\uv}$ (Fig.~\ref{fig:1DReducedSpectra}(b)),
shows a sort of power-law like behaviour at large scales. This is less
extended than the magnetic field's, so that it is not possible to
clearly distinguish whether its spectral index is closer to $-5/3$ or
$-3/2$. At scales corresponding to $k_\bot \di \gtrsim 1$,
$P_{\mathrm{1D},\bot}^{\uv}$ drops very rapidly, reaching the ppc
noise level at slightly smaller scales. Again, its parallel
counterpart in the interval $k_\parallel\,\di \sim [0.3,1]$ resembles
the perpendicular spectrum, revealing a power-law shape with a
spectral index significantly steeper than the magnetic field's,
close to $\sim -4/5$.

The reduced perpendicular spectrum of the electric field,
$P_{\mathrm{1D},\bot}^{\Ev}$ ( Fig.~\ref{fig:1DReducedSpectra}(c)),
follows $P_{\mathrm{1D},\bot}^{\uv}$ at large scales, being
$\uv\times\Bv_0$ the dominant contribution to the generalized
Ohm's law \citep{Franci_al_2015b}. It flattens when reaching the ion
scales, becoming the most energetic field.  Its spectral index
is $\sim -0.8$ at sub-ion scales (see the reference dashed line),
consistently with the fact that the dominant contributions are the
ones coming from the Hall term and from the electron pression gradient
term \citep{Franci_al_2015b}. Note
that the electric field is the most affected field by numerical noise
at small scales (susceptible to either finite-difference derivative
approximations and ppc noise effects), thus the noise is dominant
already at $k_\bot\,\di \gtrsim 1$ in the less powerful
$P_{\mathrm{1D},\parallel}^{\Ev}$.

The reduced spectra of the density fluctuations,
$P_{\mathrm{1D},\bot}^{n}$ and $P_{\mathrm{1D},\parallel}^{n}$, are
shown in Fig.~\ref{fig:1DReducedSpectra}(d). The former is almost flat
at intermediate scales, with a slope of the order of $\sim -0.7$,
although it seems to be slightly steeper at large scales. A transition
is clearly observed around ion scales, followed by a power law with a
spectral index of $\sim -2.8$.
As for the others fields, $P_{\mathrm{1D},\parallel}^{n}$ exhibits the
same profile as its perpendicular counterpart, shifted toward larger
$k$-vectors, with the sizeable difference that the parallel transfer
of energy is not negligible at all scales larger than the injection
scales, representing instead the dominant contribution up to
$k_\parallel\,\di \sim 0.4$.

A comprehensive overview of the 1D omnidirectional filtered power
spectra of all fields is provided in Fig.~\ref{fig:1DOverview}.  Since
they all exhibit power-law behavior but with different slopes, we also
show them compensated in the inset, in order to allow for a
quantitative evaluation of the spectral indices and to better
appreciate the agreement with solar wind observations.  They are all
qualitatively very similar to the corresponding reduced perpendicular
spectra, so we won't describe all the details again here. It is just
worth noting that due to the integration of the 3D spectrum over
shells of constant $k$, the power of fluctuations at small wavevector
in $P_{\mathrm{1D}}$ is slightly larger for all fields. This makes the
spectral index of the ion bulk velocity spectrum be closer to $-5/3$
(see the inset). Moreover, the density spectrum resembles a sort of
triple power-law behavior, with slopes $\sim -1., \sim -0.7, \sim
-2.8$ in the ranges $0.2 \lesssim k \di \lesssim 0.6$, $0.6 \lesssim k
\di \lesssim 2$, and $2 \lesssim k \di \lesssim 8$, respectively,
which is suggestive of solar wind observations.  The three ranges are
too small here to firmly infer the presence of three different power
laws and anyway the spectral indices do not correspond to the observed 
values of $\sim -1.8, \sim -1.1, \sim -2.8$
\citep{Safrankova_al_2015}. However, interestingly, the density
spectral behavior in an earlier phase of the evolution (not shown here) is very
similar to the one observed in the solar wind.  For example, at $t = 80$,
the three spectral index are compatible with $\sim -5/3, \sim -1.1$,
and $\sim -2.8$.  Finally, Fig.~\ref{fig:1DOverview} cleary shows how
equipartition between magnetic and density fluctuations is achieved at
kinetic scales, consistently with observations~\citep{Chen_al_2013}.
A more quantitative comparison between $P_{\mathrm{1D}}$ and
$P_{\mathrm{1D},\bot}$ will be later shown in
Fig.~\ref{fig:comparison3Dvs2D}.

\subsection{Comparison with the 2D case}
\label{subsec:comparison}

\begin{figure}[t]
\begin{center}
\includegraphics[width=0.9\linewidth]{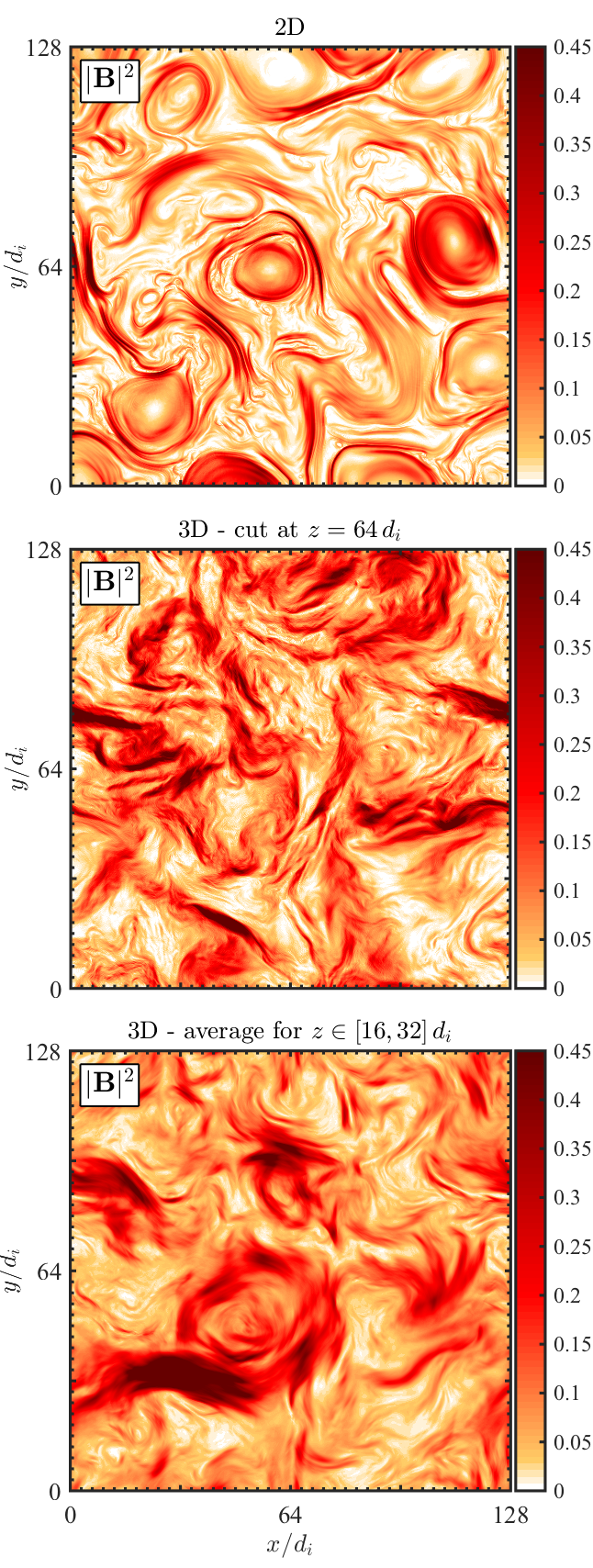}
\caption{Comparison of the magnetic structures in real space between a
  previous 2D HPIC simulation ~\citep{Franci_al_2015a} and the present
  3D one. For the former, we show a subgrid (top panel) with
  the same dimension of the full 3D box, while for the latter we show
  both a cut on a $(x,y)$ plane perpendicular to the mean magnetic
  field (middle panel) and an average along the $z$ direction over a
  length of the order of $L_{\rm box}/8 = 16\,\di$ (bottom panel).}
\label{fig:real3Dvs2D}
\end{center}
\end{figure}
\begin{figure}[t]
\begin{center}
\includegraphics[width=\linewidth]{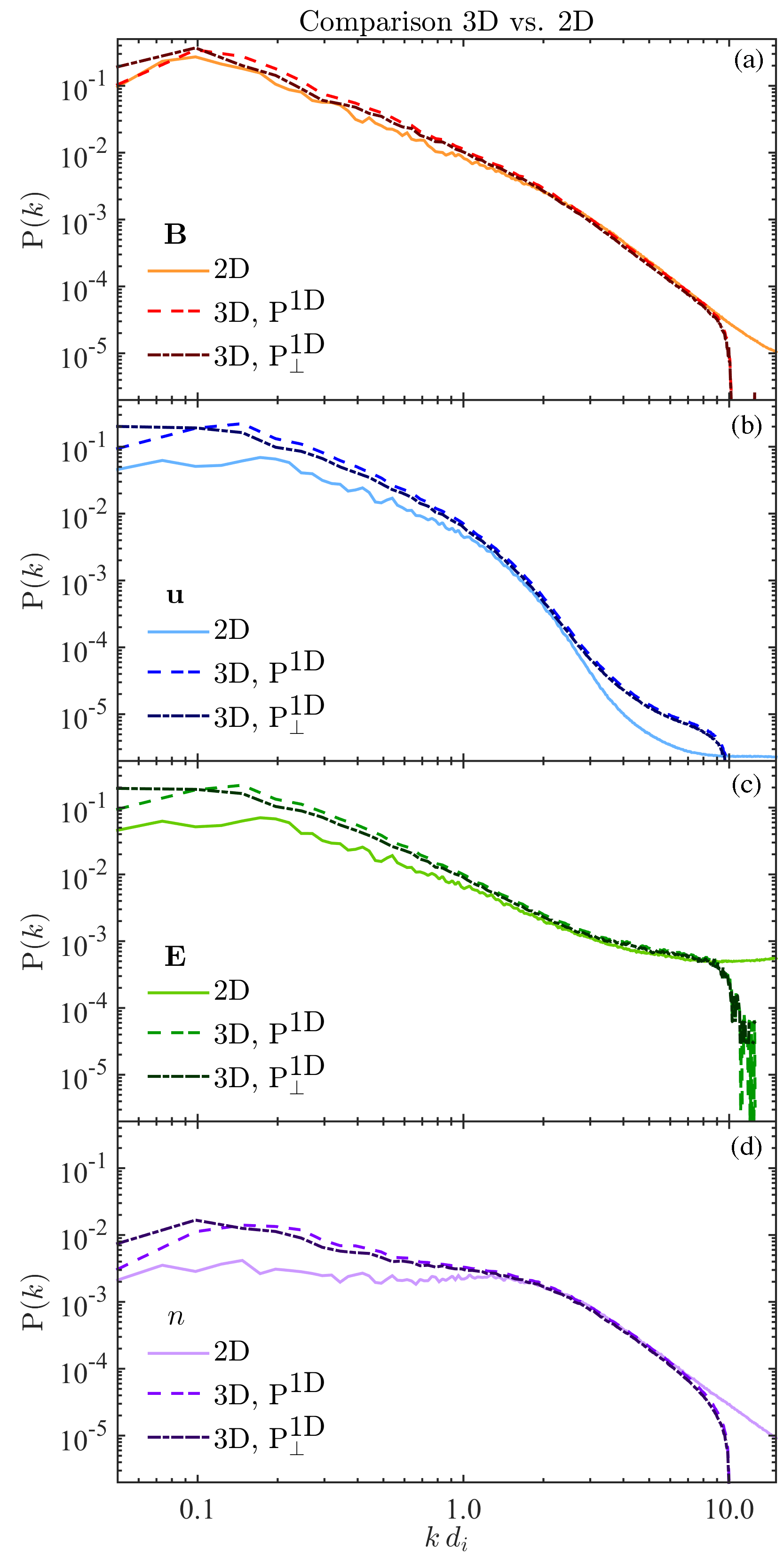}
\caption{Comparison between the spectral properties of the
  2D and the 3D simulations: magnetic (panel (a)), ion bulk velocity
  (b), electric (c), and density (d) fluctuations. For the 3D case,
  we show both the reduced perpendicular spectra and the 1D 
  omnidirectional spectra.}
\label{fig:comparison3Dvs2D}
\end{center}
\end{figure}

\begin{figure}[t]
\begin{center}
\includegraphics[width=\linewidth]{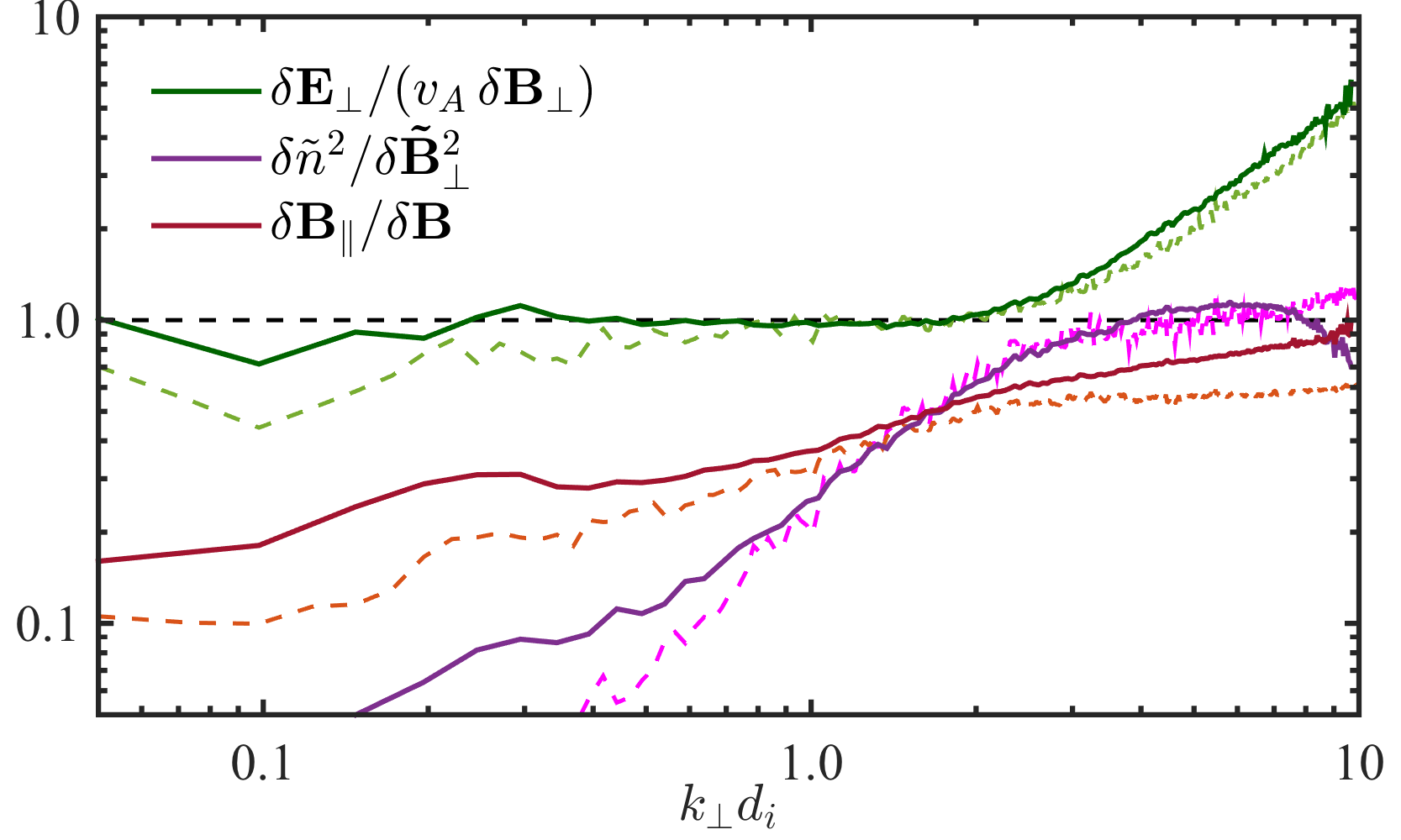}
\caption{Comparison between the spectral properties of the
  2D and the 3D simulations: ratio of the perpendicular electric field to
  the perpendicular magnetic field (green), ratio of the normalized
  density to the perpendicular magnetic fluctuations (purple), and 
  magnetic compressibility (red).}
\label{fig:ratios}
\end{center}
\end{figure}

In this section, we directly compare the results from the present 3D
simulation with those of the 2D HPIC simulation already discussed in
\citep{Franci_al_2015a,Franci_al_2015b}.  The initial conditions are
quite similar, since they are both initialized with a mean magnetic
field and the same kind of fluctuations.  These are linearly polarized
Alfv\'eninc fluctuations, which are injected at approximately the same
scales, $k\,\di \lesssim 0.25$ and $k\,\di \lesssim 0.28$, in 3D and
2D respectively, and reach a comparable level at the peak of the
turbulent activity, $\Bv^\textrm{rms} \sim 0.3$ and $\sim 0.25$. The
ion and electron plasma beta are also the same, $\beta_i = \beta_e =
0.5$. The only main difference is the resolution, since the same
accuracy is currently not feasible in 3D, due to computational
limitations: the 2D simulation have a better spatial resolution ($\Delta x =
\Delta y = \di/8$), a larger extension ($256 \di$) and a higher number
of ppc ($8000$), allowing to cover a more extended, although similar,
range in $k$-vectors, i.e., $k\,\di =[0.025,25]$ in 2D and
$[0.05,12.5]$ in 3D.

In Fig.~\ref{fig:real3Dvs2D}, we show the magnetic field
structures in real space.  In the top panel, we report $|\Bv|^2$ in a
subgrid of the 2D simualation, with the same size of the whole 3D
domain. This is compared with a 2D perpendicular 
cut of the 3D box taken at $z = 64\,\di$ (middle panel) and
with an average of the 3D box along the $z$-direction over a length of
$L_{\mathrm{box}}/8 = 16\,\di$ (bottom). In the 2D case, clear close
vortices are present, together with more elongated and filamentary
structures and small regions of strong gradients with width of the
order of $\di$. Conversely, the 3D case looks quite different when a
single perpendicular plane is selected, since circlelike structures
are completely absent. The picture changes when averaging along the
direction of the mean field, over about a correlation length in the
parallel direction (bottom panel).  Now structures resembling the
shape of vortices emerge, with a comparable size to the 2D case, and
the same holds for thin elongated regions with strong gradients,
resembling ion-scale current sheets. This suggests that the magnetic
sctructures have the shape of flux tubes oriented in the
$z$-direction, and current structures have the shape of sheets, with a
width and thickness of the same order of the 2D case. All of these
structures are modulated in the direction of the mean magnetic field,
similarly to the magnetic field lines represented in
Fig.~\ref{fig:3Dcubes}.

A direct comparison of the spectral properties between the 3D and the
2D cases is reported in Fig.~\ref{fig:comparison3Dvs2D}, where we show
the spectra of magnetic (panel (a)), velocity (b), electric (c), and
density (d) fluctuations. For the 3D case, we report both
$P_\mathrm{1D}$ (dashed lines) and $P_\mathrm{1D,\bot}$ (dot dashed).
We do not filter the 2D spectra since the energy accumulation and ppc
noise become important at small scales that are not resolved in the
3D run.

The magnetic field spectra are very close to each other, following
basically the same power laws both at fluid scales, where the 3D
simulation shows a higher power in agreement with the higher rms, and
in the sub-ion range, where they exactly overlap.  Moreover, the scale
of the transition between the two regimes is unchanged. 

The spectra of all the other fields look essentially the same in the
kinetic range (except for the higher noise level of the 3D
simulation).
Some differences are instead observed at larger (fluid) scales, e.g.,
a different level of fluctuations in the density and in the ion bulk
velocity (and, consequently, in the electric field). These could be
related to the higher level of magnetic
fluctuations injected as initial conditions in 3D, which for example
produces a higher level of compressible fluctuations at the largest
scales, or to a strong Alfv\'enic coupling between the ion bulk
velocity and magnetic fluctuations, parallel to the mean field at
$t=0$, which forces the velocity spectrum to be more coupled to the
magnetic field's at large scales. 

In Fig.~\ref{fig:ratios}, we further compare the three non-dimensional
ratios already shown for the 2D simulation in Fig. 7 of
\citealt{Franci_al_2015a}: the ratio between the perpendicular
electric fluctuations (normalized by the Alfv\'en speed) and the
perpendicular magnetic fluctuations, 
\begin{equation}
R_{E_{\bot}B_{\bot}}=\frac{\delta E_\perp / v_A}{\delta B_\perp},
\label{REB}
\end{equation}
the magnetic compressibility, 
\begin{equation}
R_{B_{\|}B}=\frac{\delta B_\parallel}{\delta B},
\label{RBB}
\end{equation}
and the ratio between the normalized energy of density and
perpendicular magnetic fluctuations,
\begin{equation}
R_{n B_{\bot}} = \frac{\delta\tilde{n}^2}{\delta\tilde{B}^2_\bot}
=\frac{\sqrt{\Gamma}\delta n/n_o}
{\delta B_\bot/B_0},
\label{RNB}
\end{equation}
where $\Gamma$ = $3/4$ for $\beta_i = 0.5$ and $T_e = T_i$, as in
the present case
\citep{Schekochihin_al_2009,Boldyrev_al_2013}.
The ratios for the 3D case are drawn with solid lines, while the
2D case is represented by lighter dashed lines.

The electric to magnetic field ratio (green) shows three
different regions.  At the injection scales, $k\di \lesssim 0.25$, it
is smaller than 1, the electric field being coupled to the velocity
field via the MHD term in the generalized Ohm's law (see discussion of
Fig.~\ref{fig:1DReducedSpectra}). The MHD-range value of
$R_{E_{\bot}B_{\bot}}$ is larger in 3D with respect to 2D because of the
stronger coupling between the velocity and the magnetic fluctuations
(the residual energy, i.e., the excess of magnetic energy over kinetic
energy, is smaller and, concurrently, the spectral index of the
velocity spectrum is closer to $-5/3$, like the magnetic field's,
rather than to $-3/2$, as in 2D).  At intermediate scales, before the
transition to the kinetic regime ($0.25\lesssim k\di\lesssim2$),
$R_{E_{\bot}B_{\bot}}$ is exactly 1, the electric field being coupled
to the magnetic fluctuations.  As seen in
Fig.~\ref{fig:comparison3Dvs2D}(a), the magnetic field spectra are
almost indistinguishable in the two simulations, meaning that the
coupling is more effective in 3D than in 2D.  At wavenumbers in the
kinetic range ($k\di\gtrsim2$) the two simulations have a negligible
difference and $R_{E_{\bot}B_{\bot}}$ increases with the same (linear)
scaling, as a direct consequence of the Ohm's law, and consistently
with observed frequency spectra in the solar wind frame
\citep{Bale_al_2005,Matteini_al_2017}.

The magnetic compressibility (red) is quite small in the MHD range and
increases toward smaller and smaller scales.  Thus, at the injection
scales, the magnetic fluctuations are mainly perpendicular to the mean
field, as imposed by our initial conditions, but they tend to become
more isotropic at ion and sub-ion scales.  The qualitative trend of
$R_{B_{\|}B}$ from fluid to kinetic scales is similar with respect to
the 2D case. However, its value is sligthly larger in 3D in the whole
range of scales, of a factor $\sim 1.5$, which is compatible with the
different initital level of fluctuations ($\Bv^{\rm rms} = 0.38$
vs. $0.24$). Note that in 2D the ratio $R_{B_{\|}B}$ reaches a plateau 
for $k\di\gtrsim2$, corresponding to component isotropy,
while in the 3D case it increases all the way down to the smallest
scales. We have checked that this is not an effect of the filtering
procedure.

Finally, the density to magnetic ratio (purple) is very small at MHD
scales, although larger in 3D than in 2D. In this range, the density
spectrum in 3D is dominated by parallel wavevectors, and it is more
energetic than in the 2D case (compare
fig.~\ref{fig:1DReducedSpectra}(d) and
fig.~\ref{fig:comparison3Dvs2D}(d)).  The difference could thus be
attributed to the parallel scales becoming accessible only in 3D.  For
$k \di \gtrsim 1$, $R_{n B_{\bot}}$ is almost the same in the two
cases, with the 3D one showing a decrease around $k\di\sim8$, only due
to the filtering procedure.  A ratio $0.75$ is predicted for kinetic
Alfv\'enic fluctuations with $\beta\sim1$
\citep{Schekochihin_al_2009,Boldyrev_al_2013}, and is observed on
average in solar wind data \citep{Chen_al_2013}.  In our 2D and 3D
runs an approximate plateau of level 1 is found, indicating that even
in 3D the perpendicular fluctuations can be reasonably described by
kinetic Alfv\'en fluctuations in an intermediate-beta case.

\section{Discussion and conclusions}
\label{sec:conclusions}

We have presented results from a large-size hybrid-kinetic 3D
simulation of freely decaying turbulence in presence of a mean
magnetic field.  The high resolution adopted in terms of both number
of grid points ($512^3$) and particle-per-cell ($2048$) allows the
simulation to develop a turbulent cascade spanning both the MHD and
the kinetic ranges of scales. As a consequence, we obtain remarkably
stable and well-defined power spectra of the magnetic, ion bulk
velocity, electric and density flucutations, covering two full decades
in $k_\bot$ and slightly more than a full decade in $k_\parallel$.

The main results of the present work are that: (i) the turbulent
cascade mainly develops in the direction perpendicular to the mean
magnetic field, so that a strong spectral anisotropy is achieved,
despite the isotropic initial conditions; (ii) the 1D omnidirectional
spectra (as well as the 1D reduced perpendicular spectra) of all
fields exhibit power-law behavior spanning both the MHD and the
kinetic range, with spectral indices that are in remarkable agreement
with solar wind observations; (iii) the comparison between the present
3D simulation and a previous 2D simulation with similar
parameters shows that the spectral behavior of all fields near and
below the ion characteristic scales is not affected by a
reduced 2D geometry.
 
The 3D spectrum of the magnetic fluctuations shows that turbulence
develops a strong anisotropy, with more power in the perpendicular
wavevectors. In particular, the anisotropy is observed to be
scale-dependent at large MHD scales. On the contrary, it becomes
scale-independent at small kinetic scales.  This last result is also
confirmed by the 1D reduced parallel and perpendicular spectra: for
all the fields, the former exhibits the same power law of the latter,
just shifted toward larger scales.  At even larger scales, the
parallel spectra do not show any power-law behavior, possibly because
the physical box size along $B_0$ is too short to accomodate the long
parallel wavelengths expected in a strong-turbulence regime.  We
expect anisotropy to change when analysed in a reference frame
attached to the local mean field.  A more quantitative analysis and an
articulated discussion about the spectral anisotropy is beyond the 
scope of this paper and will be the subject of future work.

The large physical extent in the perpendicular directions allows us to
recover power-law scaling in both the MHD and the kinetic range,
covering simultaneously two full decades in wavenumbers and the
transition between the two regimes.  The 1D omnidirectional
spectra recover several spectral indices found in larger
2D HPIC simulations with similar initial conditions and plasma
parameters~\citep{Franci_al_2015a,Franci_al_2015b}.

In particular, the magnetic field spectrum has the same properties as in 2D
HPIC simulations at both MHD and kinetic scales: a double power-law
behavior with a Kolmogorov-like index of $-5/3$ at low $k$s, followed
by a steeper spectrum with index $\sim-3$ in the kinetic range.  The
transition (break) occurs around ion scales, $k \,\di\sim 2$, just as 
in the 2D case with the same ion plasma beta. This support our previous
2D numerical study in which we provided numerical evidence that the
plasma beta controls the position of the break~\citep{Franci_al_2016b}.

The ion bulk velocity, electric field, and density spectra are again
similar to the corresponding 2D ones, but only at kinetic scales.  The
ion bulk velocity spectrum drops dramatically, with a trend that can
be approximated as a power law with a steep slope of $\sim-4.5$.  The
electric field power flattens with a power-law index $\sim -0.8$,
becoming the dominating type of fluctuations.  The density spectrum
steepens and reaches a sort of equipartition with the magnetic energy
at sub-ion scales. Its power-law index, $\sim -2.8$, is close
but not identical to the one of the magnetic field spectrum.
All these spectral properties in agreement with the 2D case indicate
that, because of the large anisotropy, the coupling between the
electric field fluctuations, density fluctuations, and the parallel
component of magnetic fluctuations found in 2D at kinetic scales
\citep{Franci_al_2015a} continues to hold in a full 3D geometry.
Note that all the above spectral indices are also in agreement with
 solar wind observations.

Al large scales, on the contrary, the velocity, electric field, and
density spectra have noticeable differences with respect to the 2D case
and to observations in the solar wind.  

The velocity fluctuations have a power-law index closer to $-5/3$ than to
$-3/2$, as found instead in 2D simualtions and in observations.  In our 3D
run, they look strongly coupled to the magnetic fluctuations.  Note that
the indices $-3/2$ and $-5/3$ are found for the velocity and magnetic
fluctuations in 2D HPIC simulations with an out-of-plane mean field
\citep{Franci_al_2015a,Franci_al_2015b}, and in weakly compressible 3D
MHD simulation with no mean field \citep{Grappin_al_2016}.  This
suggests that only a small Alfv\'enic coupling is achieved in the
solar wind, possibly because of the value $B^\mathrm{rms}/B_0\sim 1$
at hour scales and of the small, but non-negligble, compressibility of
velocity fluctuations.

The electric field is strongly coupled to the velocity spectrum at
large scales, consistently with the generalized Ohm's law
\citep{Franci_al_2015b} and solar wind observations. Moreover, it
keeps following the magnetic field spectrum as the velocity
fluctuations start dropping just above the ion scales.

The density spectrum is more energetic and steeper than in the 2D case
at large scales. This could be due to the fact that the energy in
parallel wavevectors is larger than in the perpendicular ones, and
that parallel scales are clearly absent in 2D. Here, instead, this
allows for a larger compressibility in the MHD range. Moreover, the
higher level of $B^{\rm rms}$ with respect to the 2D case activates a
sufficient power in the density fluctuations, so that the hint of a
short decrease (possibly a cascade) is observed at MHD scales. This is
consistent with the behavior observed in 2D simulations with larger
$B^{\rm rms}$ (not shown here). Note that the
density spectrum resembles a peculiar triple power-law behavior. This
is much more evident and clear during the development of the turbulent
cascade, when the three observed spectral indices are also reproduced,
than during the fully-developed quasi-stationary state.

Accordingly to the spectral anisotropy, large-scale structures and
currents in real space are preferentially aligned to the mean magnetic
field, with gradients being more developed in the perpendicular plane.
When averaging over a parallel correlation length, such structures are
isotropic in the perpendicular plane and acquire the vortex-like shape
characteristic of 2D runs.  In a perpendicular cut, small-scale
currents have roughly the same thickness and width found in a 2D
geometry. Their aspect ratio is approximately conserved after
averaging in the parallel direction, indicating that currents are
sheet like stuctrures with a long-wavelength parallel modulation and a
weak twist.

The agreement between the 2D and the 3D case is important both from a
pratical and a physical point of view. On the one hand, it validates
the use of 2D simulations for all those cases where the study of
spectral properties is involved and the use of a large collection of
simulations with different values of parameters is required, e.g., for
convergence studies (\citep[e.g.][]{Franci_al_2015b}), of for parameter
studies (\citep[e.g.][]{Franci_al_2016b}).  On the other hand, it suggest
that the dominant process(es) responsible for the ion-scale spectral
break in the magnetic field and for the kinetic-scale turbulent
cascade are likely not inhibited in a reduced 2D geometry.  In
particular, the fact that the scale of the break is completely unmoved
passing from a 2D evolution to 3D when the same ion plasma beta is set,
supports our previous finding that beta is likely the only main
parameter controlling such scale. Moreover, the fact that the width of
the 2D current sheets in the 3D simulation seems to be of the same
order of the 1D current sheets in the 2D simulation (cf.  the bottom
panel of Fig.~\ref{fig:real3Dvs2D}) might support, although only
qualitatively at this level, the idea that the disruption of current
structures via magnetic reconnection may be the main responsible for
the break and the onset of the sub-ion scale cascade even in more
realistic 3D turbulence.
Although the complex shape of magnetic and current structures makes
the identification of reconnection sites in 3D much more difficult
than in 2D, a preliminary analysis (not presented here) shows a  
correspondance between the first peak in the maximum of the current
and the development of the kinetic power law is observed, in agreement
with the findings of~\citet{Franci_al_2017} in 2D.
A more quantitative and detailed analysis of the current structures
would be necessary in order to confirm this scenario and will be the
subject of future work.

\section{Acknowledgments} 

The authors wish to acknowledge valuable discussions with E. Papini, 
C. H. K. Chen., S. S. Cerri, F. Califano, D. Burgess, and O. Alexandrova.
LF is funded by Fondazione Cassa di Risparmio di Firenze through
the project ``Giovani Ricercatori Protagonisti''. PH acknowledges GACR
grant 15-10057S. We acknowledge PRACE for awarding us access to
resource Cartesius based in the Netherlands at SURFsara through the
DECI-13 (Distributed European Computing Initiative) call (project
HybTurb3D), and the CINECA award under the ISCRA initiative, for the
availability of high performance computing resources and support
(grants HP10C877C4 and HP10BUUOJM). Data deposit was provided by the
EUDAT infrastructure (\url{https://www.eudat.eu/}).

\bibliographystyle{apj-eid}

\end{document}